\documentclass{article}
\usepackage[T1]{fontenc}
\usepackage[latin1]{inputenc}
\usepackage[dvips]{graphicx}

\makeatletter

\providecommand{\LyX}{L\kern-.1667em\lower.25em\hbox{Y}\kern-.125emX\@}

\makeatother

\begin{document}

\title{An alternate smearing method for Wilson loops in lattice QCD}

\author{F. Okiharu$^1$, R.M. Woloshyn$^2$\\
$^1$Department of Physics, Faculty of Science and Technology,\\
Nihon University, 1-8-14 Kanda-Surugadai, Tokyo 1018308, Japan\\
$^2$TRIUMF, 4004 Wesbrook Mall, Vancouver BC, Canada V6T 2A3}

\date{ }
\maketitle
\begin{abstract}
A gauge field link smearing method developed for calculations with staggered
fermions, namely the use of unitarized fat7 links, is applied to mesonic and
baryonic Wilson loop calculations. This method is found to be very effective
for reducing statistical fluctuations for large Wilson loops. Examination of
chromo-electric field distributions shows that self-interactions of the static
sources are reduced when unitarized fat7 smearing is used but long-distance
inter-quark effects are unchanged.
\end{abstract}

\section{Introduction}

Wilson loops are used in many applications of lattice QCD, for example, in the
study of the static quark potential, string breaking and the chromo-electric
and chromo-magnetic field distributions. These calculations can be very challenging
for Monte-Carlo simulations because the Wilson loops decrease exponentially
with the size of the loop (area law) and rapidly become submerged in statistical
fluctuations. Usually some method that will enhance the signal and suppress
the noise has to be used. The multihit method\cite{parisi, ForRoi} for 
reducing the variance of the
signal and link smearing which enhances the ground state signal are well known.
As well there are special techniques like abelian projection which preserves
the essential long distance physics while removing extraneous random 
fluctuations (see \cite{ichie} for a recent application). 

The above methods work because they smooth out the gauge fields at distances
less than some physically relevant length scale. In other words, the effects
of hard gluons, gluons with momenta of the order of the cut-off, are suppressed
by some kind of smearing. There is another context in which the suppression
of hard gluons is also very important. With staggered fermions on a lattice
the so-called species doubling occurs. Multiple copies of staggered fermions
(now usually called tastes) exist. During the past few years it has been understood
that taste symmetry breaking, which is a finite lattice spacing artefact, is
due to hard gluon interactions\cite{lepage,orginos}. 
Improved staggered fermion actions utilizing
fat gauge field links have been developed and have been shown to greatly reduce
taste-symmetry breaking\cite{orginos,blum,hasen}. 

It is natural to ask if the methods developed for staggered fermions are also
effective in Wilson loop calculations. Indeed it has already been shown that
hypercubic blocking, which is used in the so-called HYP action, can reduce the
error in static quark-antiquark potential calculations\cite{hasen}. 

Recently, further improved staggered fermion actions have been developed\cite{folla}. 
These are based on the idea of unitarized fat7 smearing first discussed by
Lee\cite{wlee}. 
In the most common smearing method, sometimes
called APE smearing\cite{ape}, 
one adds to the target link its three link staples with
some coefficient and then projects this combination of links back to \( SU(3) \).
This process is repeated. The relative weighting of the target link and staples
and the number of smearing steps are determined essentially by trial and error
for different applications. In fat7 smearing one adds not just the 3-link staples
but also 5-link and 7 link staples which extend in mutually orthogonal directions,
orthogonal to the direction of the target link (see Ref. \cite{orginos}). 
The weighting of
the staples is fixed by requirements of action improvement. What has been realized
recently is that unitarization or projection to \( SU(3) \) is very beneficial
for suppression of taste-changing interactions. A nice feature of the unitarized
fat7 (Ufat7) scheme is that it is easy implement compared to hypercubic
blocking\cite{hasen}.
It is shown here that it is also very effective for Wilson loop calculations.

The fat7 smearing operator involves many more terms than the commonly used APE
smearing. For a single smearing of one link, fat7 takes about eight times longer
than APE smearing. In typical applications of APE smearing many tens of 
smearing steps are employed. However, as will be seen from our results, only 
a few Ufat7 smearing steps are needed to achieve a significant reduction of 
statistical errors. Therefore, Ufat7 smearing can be used without adversely 
impacting the overall cost of the calculation.

In Sect. 2 results for the static quark-antiquark potential are presented. Multiple
levels of Ufat7 smearing are considered. It is found that the long-distance
linearly rising behavior of the potential is preserved by smearing but the Coulomb-like
part of the potential at short distances is suppressed. The main effect of
smearing is to remove the constant term in the potential. This term is associated
with the quark self-energy. A nice way to see how this is being affected by
smearing is to look at the distribution of color field around the static sources.
This can be done by calculating the Wilson loop - plaquette correlation\cite{bali}. 
Results for chromo-electric field distributions are presented and with Ufat7 smearing
one can see the emergence of a flux tube connecting the static sources even
for a fairly small spatial separation.

The baryonic Wilson loop is considered in Sect. 3. This case is much noisier
than the mesonic loop. However, aggressive use of Ufat7 smearing is found to
reduce statistical errors considerably for the potential at large quark separations
compared to the multihit method that has been used before. As with the quark-antiquark
potential, the constant term of the potential is essentially removed after a
few levels of smearing. Examination of the chromo-electric field distribution
shows that self-interaction effects at the static quark sources are highly suppressed
by smearing and the inter-quark flux can be revealed.

\section{Mesonic Wilson loop}

The interaction energy of a static quark-antiquark pair can be determined by
taking the large time limit of a Wilson loop \( W(R,T) \) of spatial extent
\( R \)

\begin{eqnarray}
V(R,T) & = & -\ln (W(R,T)/W(R,T-1)),\\
 & \rightarrow  & V_{Q\overline{Q}}(R).
\end{eqnarray}
 Both on-axis and off -axis Wilson loops are considered and the static potential
is calculated for 24 different quark-antiquark separations up to a maximum
separation of 9 lattice units. The numerical simulations were carried out in
quenched approximation using the Wilson plaquette action at \( \beta =6.2 \)
on a \( 24^{4} \) lattice. 

In order to reduce the statistical error some noise reduction technique is commonly
used. The multihit method\cite{parisi, ForRoi} has been found to be very effective 
for static potential calculations. Essentially one integrates out the gauge links in 
the time direction,
replacing them by effective links constructed from the staples attached to the
replaced temporal link. In addition, fuzzing or smearing of the links in the
spatial segments of the loop is also carried out. This improves the overlap
with the ground state and allows the use of (1) at moderate values of T. For
the spatial links in the loop, a standard way to do this smearing is to make
the replacement\cite{ape} \begin{equation}
U_{i}(x)\rightarrow V=U_{i}(x)+\alpha \sum _{j\neq i}U_{j}(x)U_{i}(x+\widehat{j})U^{\dagger }_{j}(x+\widehat{i})
\end{equation}
 with the sum on \( j \) denoting the sum of staples in spatial directions
orthogonal to \( i \). The combination of links is then projected back to \( SU(3). \)
In this paper we use the projection method in Ref. \cite{hoek} which iteratively finds the
element \( U'\in SU(3) \)which maximizes \( ReTr(U'V^{\dagger }) \). 

In Fig. 1 the static quark-antiquark potential is plotted for a calculation
using an ensemble of only eight gauge configurations.
The circles in Fig. 1 show the potential (in lattice units) for a standard
calculation with multihit and 30 steps for spatial link smearing (3) with \( \alpha =0.15 \).
The potential obtained using Eq. (1) at \( T=8. \) The other data sets of Fig. 1
are calculations utilizing one, two or three levels of Ufat7 smearing for all
links. In addition to Ufat7 smearing, spatial links are smeared with 8 levels
of (3) which is sufficient to get good ground state overlap. The decrease of
statistical errors achieved with Ufat7 smearing is evident.

To investigate how well the long distance physics is preserved by smearing a
high statistics calculation with 1500 gauge configurations was carried out.
The potentials are shown in Fig. 2(a). The statistical errors are too small
to be plotted. In Fig. 2(b) the
different data sets are plotted shifting the points so that the value at \( R=9 \)
is the same. The long distance behavior of the potential is preserved remarkably
well.

In order to understand what Ufat7 smearing is doing it is useful to look at
color field distributions. We calculate the correlation function between the
Wilson loop \( W(R,T) \) and the plaquette operator \( P_{\mu \nu }(x,t) \)
given schematically by

\begin{equation}
\frac{<WP>}{<W>}-<P>.
\end{equation}
 Choosing a temporal plaquette \( P_{i0} \) yields the square of the chromo-electric
field in the \( i \)-th direction. In the actual calculation the plaquette
operator is symmetrized about its spatial base point, that is, the operator
\( (P_{i0}+P_{-i0})/2 \) is used\cite{bali}. The temporal position of plaquette is 
taken to be at the mid-point of the Wilson loop, \( t=T/2 \). Using spatial plaquettes
\( P_{ij} \) chromo-magnetic fields can also be probed. This was done but the
qualitative features are the same as for chromo-electric fields so no results
for chromo-magnetic fields are presented here.

Results are presented for a planar \( 6\times 6 \) Wilson loop. Although this
is not a very large loop it is large enough to illustrate the effect of Ufat7
smearing with the advantage that the statistical errors for our high statistics
run are small enough to
be omitted for the plots. Fig. 3 shows the square of the chromo-electric field
as a function of the longitudinal coordinate \( r_{||} \), the position in
the plane of the loop parallel to the spatial segments of the Wilson loop. The
field components parallel and transverse to the longitudinal axis are plotted
separately. In the coordinate system that was used, the quark and antiquark
source positions correspond to \( r_{||}=4 \) and \( r_{||}=10. \) The circles
show the results of a standard calculation, multihit noise reduction for temporal
links and standard APE-like smearing of spatial links. For the small loop used
here, the flux connecting the quark and antiquark sources is completely submerged
in the huge flux that surrounds each source. However, successive levels of Ufat7
remove these short distance self-interaction effects exposing the nascent flux
tube between the sources. This is reenforced by examining the transverse profiles
of the \( E^{2} \) distributions. These are shown in Fig. 4 and 5 for the parallel
and transverse field components. The plane of the Wilson loop corresponds to
\( r_{\bot }=5. \) After three levels of Ufat7 smearing, chromo-electric flux
outside the Wilson loop is effectively removed and the transverse profile is
essentially the same at all longitudinal positions inside the loop. This is
what is expected for confined flux that gives rise to a linearly rising potential.

\section{Baryonic Wilson loop}

The interactions of three static quarks can be obtained using a three-bladed
Wilson loop\cite{ichie,takahash,alex}
depicted in Fig. 6. The color indices associated with the three
loop segments \( S_{1},S_{2} \) and \( S_{3} \) are contracted according to
\begin{equation}
W_{3q}=\frac{1}{3!}\epsilon ^{abc}\epsilon ^{a'b'c'}S_{1}^{aa'}S_{2}^{bb'}S_{3}^{cc'}.
\end{equation}
 The quark sources are at \( r_{1},r_{2},r_{3} \) and the position of junction
point can chosen arbitrarily, at least, in principle. In practise it may be difficult
to achieve the conditions where observables are independent of the shape of
the operator used to create and annihilate the three-quark state\cite{okiha}.

To calculate the interaction energy of the three-quark system we use a source
and sink where each of the quarks lie on a different spatial axis at a distance
\( l \) from the origin\cite{takahash,alex}. 
The junction point of the links connecting the sources
is at the origin. The inter-quark separation is \( \sqrt{2}l \) and the sum
of the distances from the quark sources to the so-called Steiner point is \( \sqrt{3}l \)
. The three-quark potential is shown in Fig. 7(a) (circles) for a standard calculation
using multihit variance reduction and 40 steps of spatial link smearing. The
potential is calculated at T = 8 (see (1)) which is sufficient to have a reasonable
plateau. Also shown are the potentials with different levels of Ufat7 smearing.
The Ufat7 smeared calculations also included spatial link smearing using (3)
for 20 steps with \( \alpha =0.25. \) In Fig. 7(b) the results are plotted with
a constant shift so that the potentials agree at \( l=6. \)

The suppression of statistical fluctuations at large distance is very evident
with Ufat7 smearing. The linearly rising long distance behavior is maintained
just as in the mesonic case and the main effect is the suppression of the constant
potential. This is due to a reduction of the short-distance self-interaction
effects at the static sources. Examination of the color field distribution confirms
this effect. 

The calculation of the flux with a baryonic source is quite challenging so we
use a loop with only fairly small separations. The source for this calculation
is planar with the shape of a $\top$. The length of the crossarm of the $\top$ 
is 10 lattice
units and the length of the base is 8. The time extent of the loop is 6 with
the plaquette probing the flux distribution at time 3. As an example of the
flux distribution Fig. 8 shows \( E^{2}_{||} \) the square of the components
of the chromo-electric field in the plane of the baryonic source for different
smearings. The scale in each figure is different so that only a qualitative
comparison is possible. As Ufat7 smearing is increased the peaks at the sources
are suppressed and the flux connecting the sources becomes visible. This is
the same as what was seen in the mesonic case but it seems that more smearing
is needed in the baryonic case to achieve a similar level of source suppression. 

The behavior of the chromo-electric field is illustrated more quantitatively
in Fig. 9. The quantity  \( E^{2}_{||} \) (in lattice units) is plotted on
a line along the base of the baryonic operator. Recall that the operator has 
the shape of a $\top$. The point  \( r=12 \) corresponds to the source point at
the bottom of the $\top$ and the point  \( r=4 \) is the point where the base
meets the crossarm. One sees not only the suppression of the self-interaction of
the source but also the very significant decrease in the statistical errors
achieved through the use of Ufat7 smearing.

The baryonic loop used in the flux distribution measurement is still too small
to claim an observation of baryonic flux tubes. However with a combination of
increased statistics, more aggressive Ufat7 smearing and some tuning of the
spatial smearing it may be possible to go to loops large enough to reveal some
interesting features of the color-field flux.

\section{Summary}

With staggered fermions on a lattice, hard gluons can lead to taste changing
transitions. Recently a strategy using a unitarized fat7 smearing has been developed
to reduce these effects in hadrons simulated with staggered quarks. In this work
we investigate the use of unitarized fat7 smearing for mesonic and baryonic
Wilson loops.

It is found that unitarized fat7 smearing is very effective in reducing statistical
errors in large Wilson loops. This comes about because the smearing suppresses
short distance fluctuations associated with self-interaction of the static sources.
An indirect manifestation of this suppression is the reduction of the constant
term in the potentials as more levels of Ufat7 smearing are applied. The examination
of chromo-electric flux distributions shows that large peaks associated with
the sources in standard calculations can be removed substantially by using Ufat7
smeared links. The long distance part of potential and the flux exchanged between
sources do not seem to be disturbed by the smearing.

The fact that a method developed to suppress hard gluon interactions of staggered
quarks also works well for Wilson loops encourages us to think that relevant
physical effects have been identified and controlled. It is hoped that unitarized
fat7 smearing will find useful applications where large Wilson loops are calculated.

\noindent
{ \it Acknowledgements.} This work is supported in part by the 
Natural Sciences and Engineering Research Council of Canada.

\newpage
\section*{Figure captions}

\textbf{Fig. 1.} The static quark-antiquark potential (in lattice units) versus
inter-quark separation in a low-statistics run. A standard calculation using 
multihit variance reduction
is shown by circles. Results for one, two and three levels of Ufat7 smearing
are represented by squares, triangles and diamonds respectively. 

\medskip{}
\noindent \textbf{Fig. 2.} The static quark-antiquark potential (in lattice units) versus
inter-quark separation. A standard calculation using multihit variance reduction
is shown by circles. Results for one, two and three levels of Ufat7 smearing
are represented by squares, triangles and diamonds respectively. In (b) a constant
shift has been applied. 

\medskip{}
\noindent \textbf{Fig. 3.} The square of the chromo-electric field (in lattice
units) (a) parallel and (b) perpendicular to the spatial axis of the Wilson
loop as a function of the longitudinal coordinate \( r_{||}. \) The source positions
are \( r_{||}=4 \) and \( r_{||}=10. \)

\medskip{}
\noindent \textbf{Fig. 4.} The profile of the parallel chromo-electric field
squared distribution (in lattice units) as a function of position \( r_{\bot } \)
transverse to the spatial axis of the loop. The different panels are for different
longitudinal positions \( r_{||}=2,3,4,5,6 \) and \( 7. \) The source is at
\( r_{||}=4,\: r_{\bot }=5. \) 

\medskip{}
\noindent \textbf{Fig. 5.} The profile of the transverse chromo-electric field
squared distribution (in lattice units) as a function of position \( r_{\bot } \)
transverse to the spatial axis of the loop. The different panels are for different
longitudinal positions \( r_{||}=2,3,4,5,6 \) and \( 7. \) The source is at
\( r_{||}=4,\: r_{\bot }=5. \)

\medskip{}
\noindent \textbf{Fig. 6.} Three bladed Wilson loop used for determining the three-quark
potential.

\medskip{}
\noindent \textbf{Fig. 7.} The static three-quark potential (in lattice units)
for a source where each of the quarks lie at a distance \( l \) from the origin.
A standard calculation using multihit variance reduction is shown by circles.
Results for one, two and three levels of Ufat7 smearing are represented by squares,
triangles and diamonds respectively. In (b) a constant shift has been applied.

\medskip{}
\noindent \textbf{Fig. 8.} Surface plot showing the distribution of chromo-electric
field squared parallel to the plane of the three-quark source for (a) standard
calculation with multihit, (b) one level of Ufat7 smearing, (c) two levels of
Ufat7 smearing, (d) three levels of Ufat7 smearing.

\medskip{}
\noindent \textbf{Fig. 9.} The square of the chromo-electric field 
(in lattice units) in the plane of the baryonic operator. The coordinate
\( r \) is on a line along the base of the operator with the source at
\( r=12. \)

\pagestyle{empty}
\vspace*{\fill}
\begin{center}
\begin{figure}
\includegraphics[angle=180]{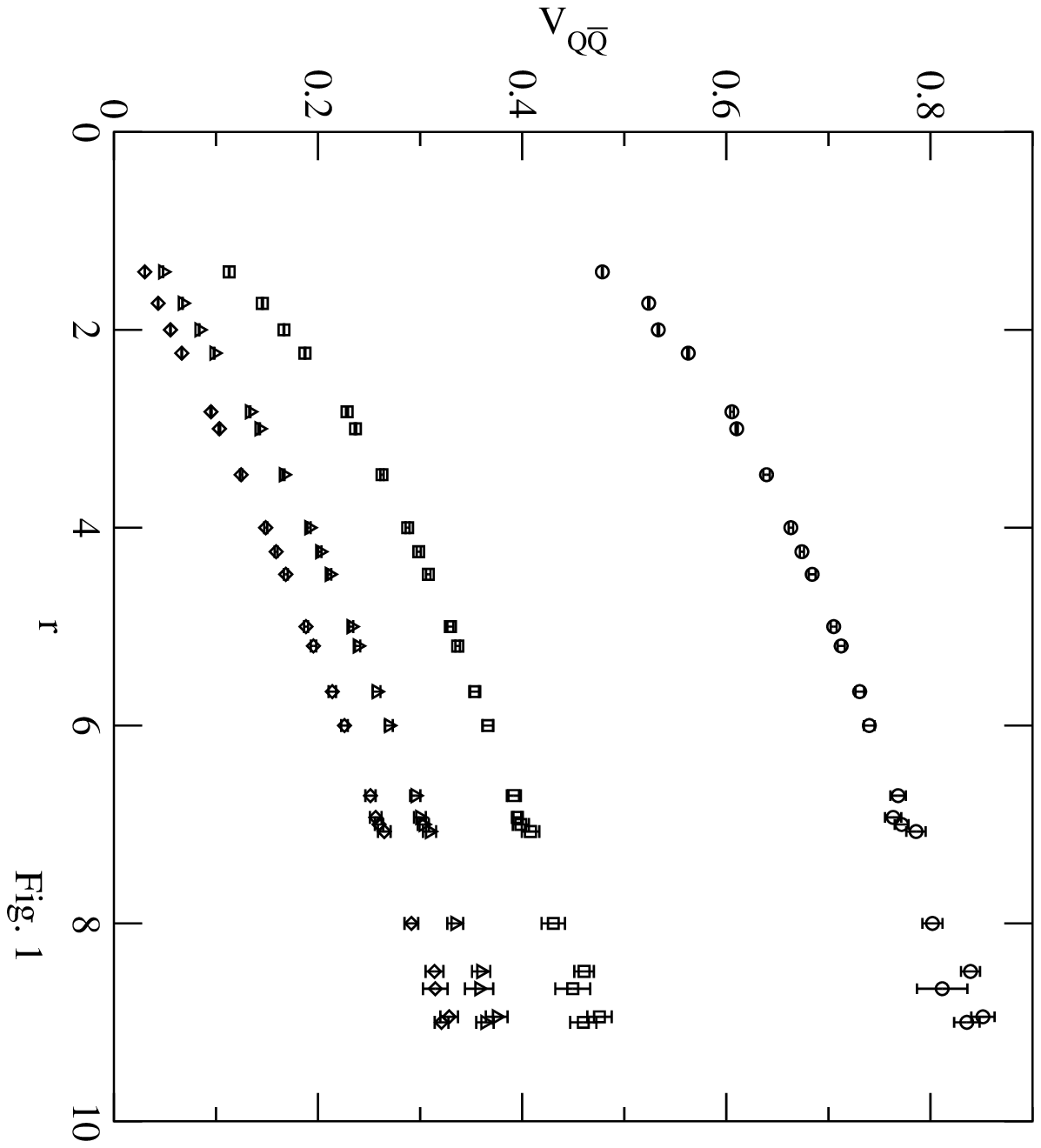}
\end{figure}
\end{center}
\vspace*{\fill}

\pagestyle{empty}
\vspace*{\fill}
\begin{center}
\begin{figure}
\includegraphics{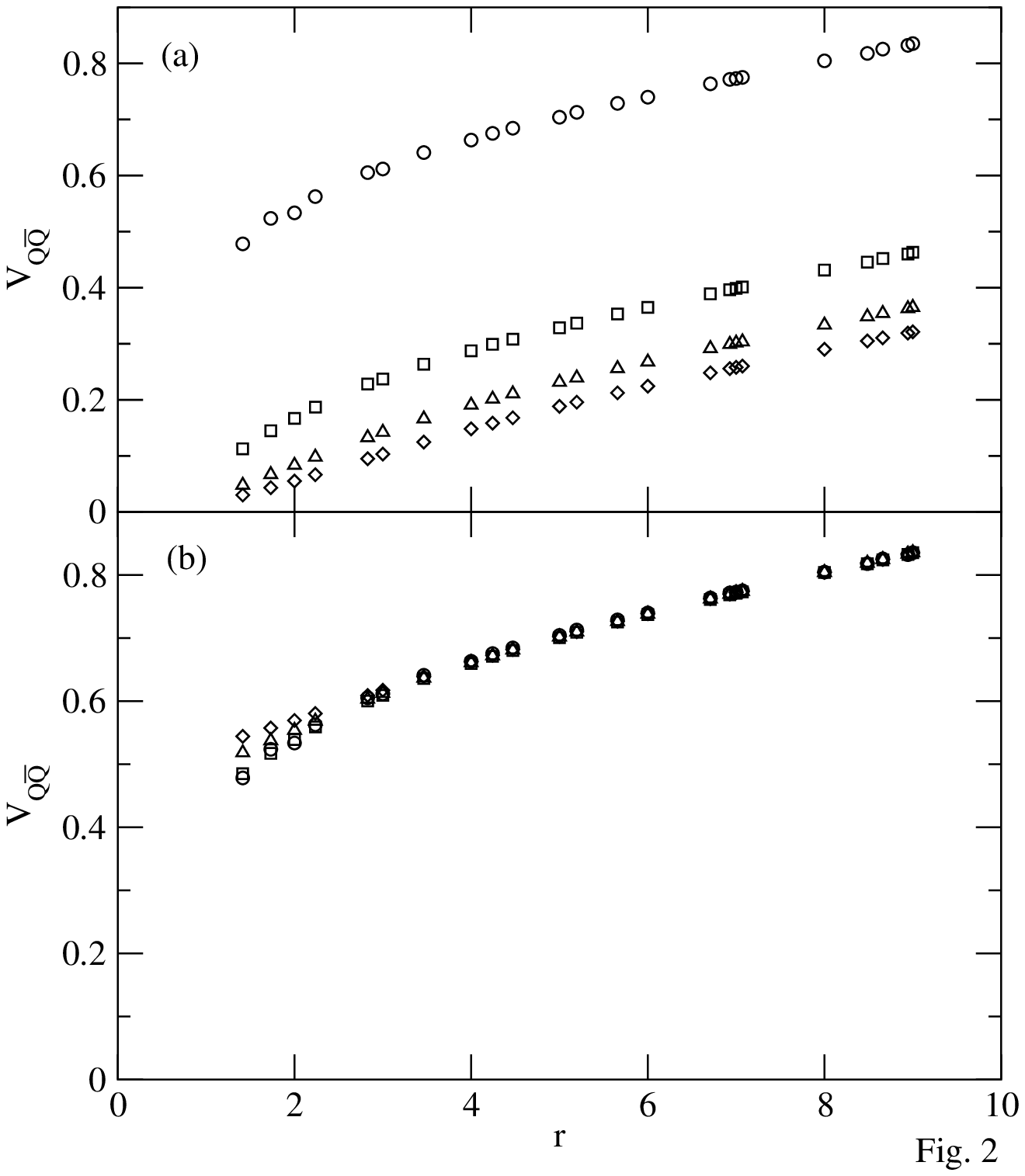}
\end{figure}
\end{center}
\vspace*{\fill}

\pagestyle{empty}
\vspace*{\fill}
\begin{center}
\begin{figure}
\includegraphics{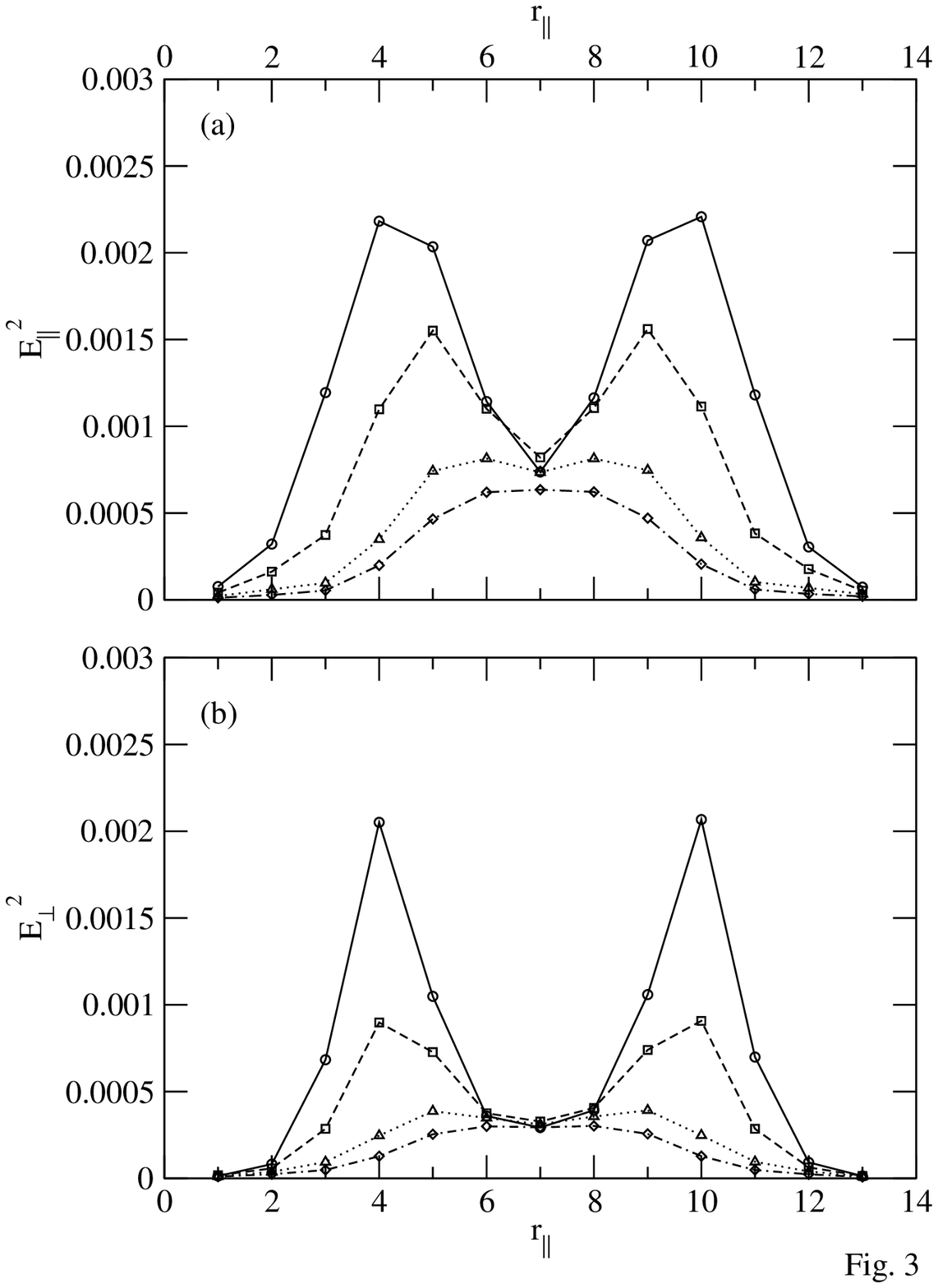}
\end{figure}
\end{center}
\vspace*{\fill}

\pagestyle{empty}
\begin{figure}
\includegraphics[angle=180]{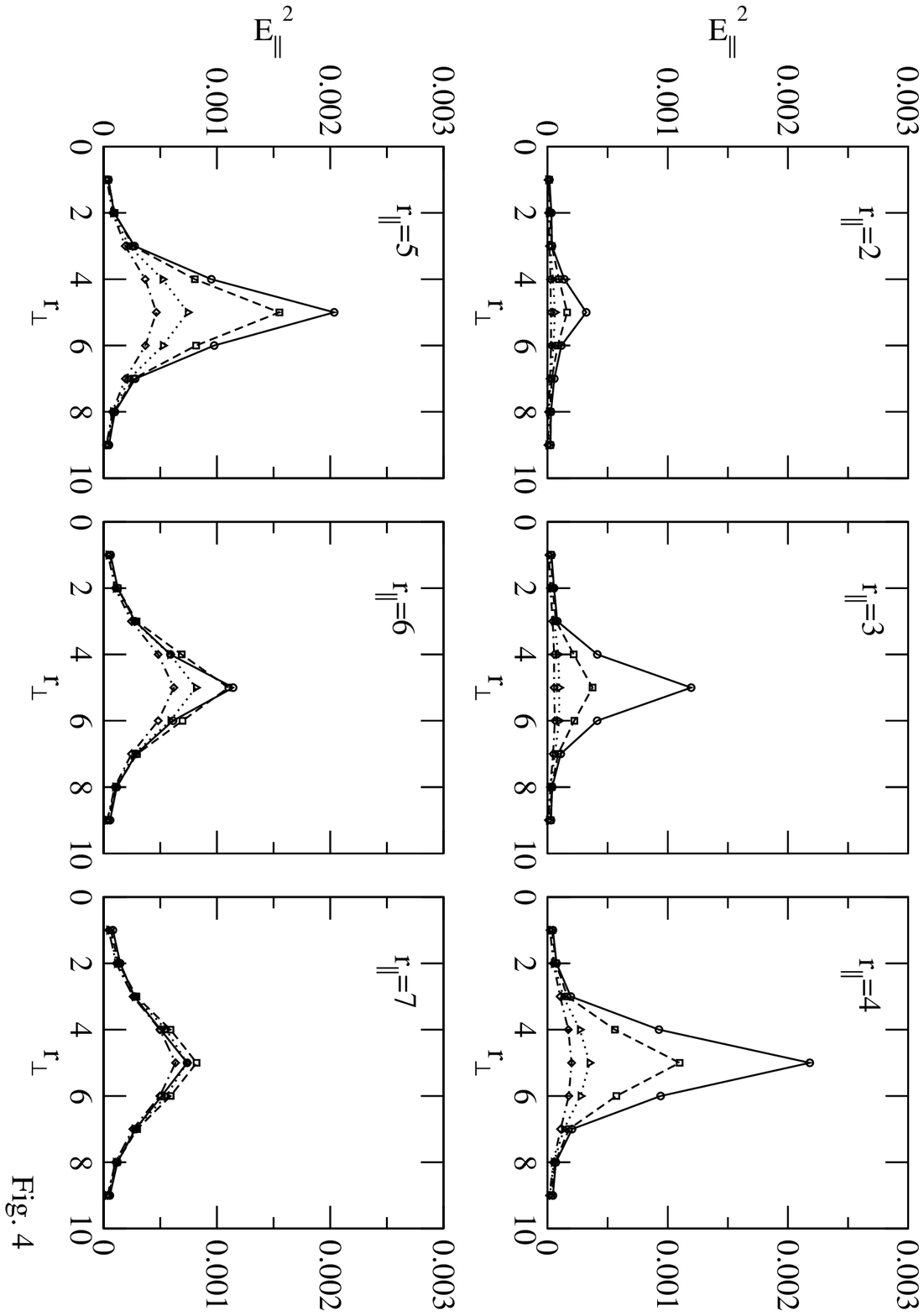}
\end{figure}

\pagestyle{empty}
\begin{figure}
\includegraphics[angle=180]{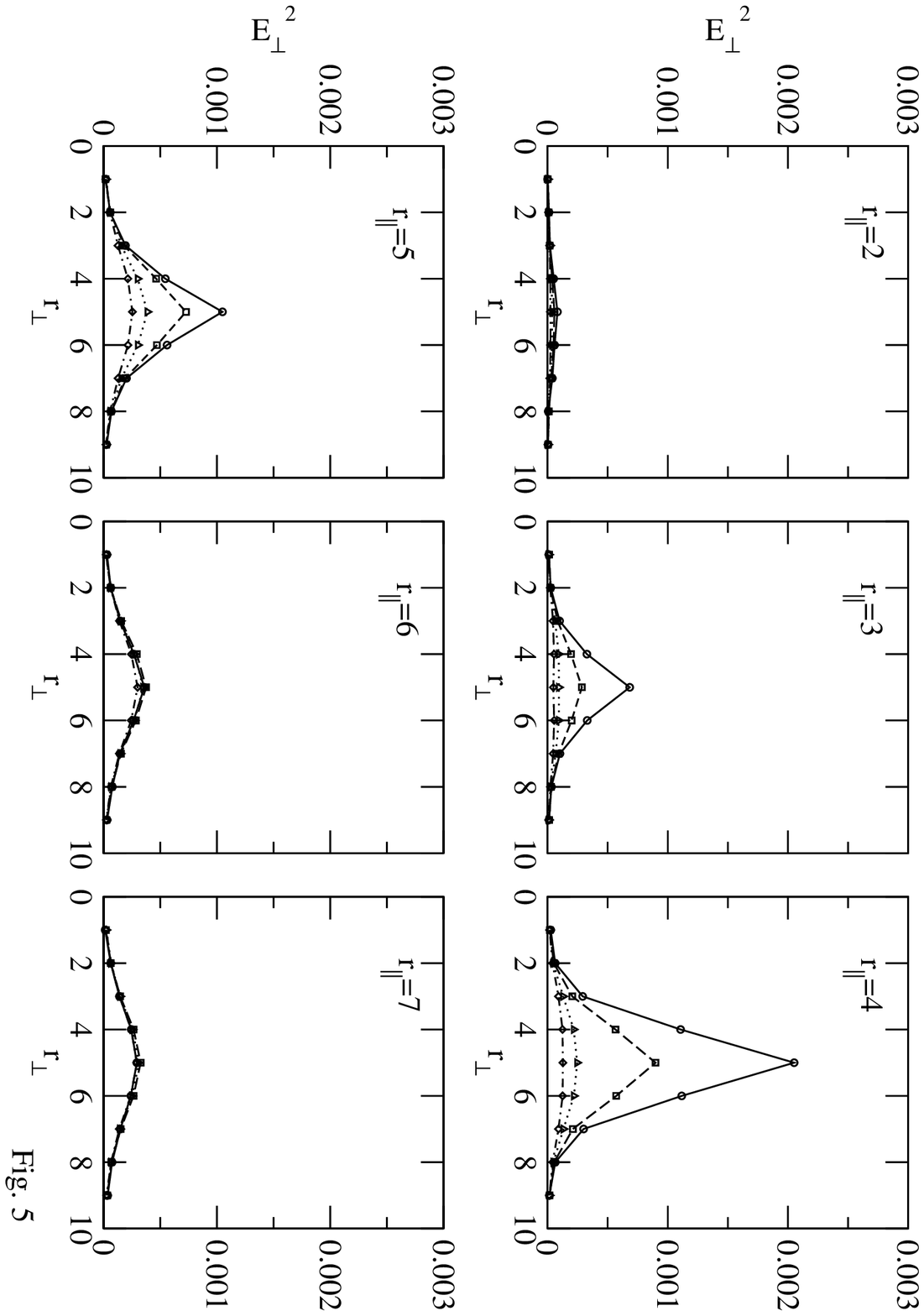}
\end{figure}

\pagestyle{empty}
\begin{figure}
\includegraphics{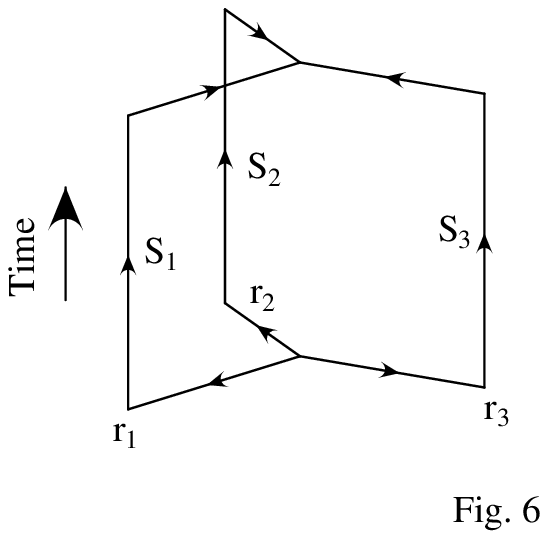}
\end{figure}

\pagestyle{empty}
\vspace*{\fill}
\begin{center}
\begin{figure}
\includegraphics{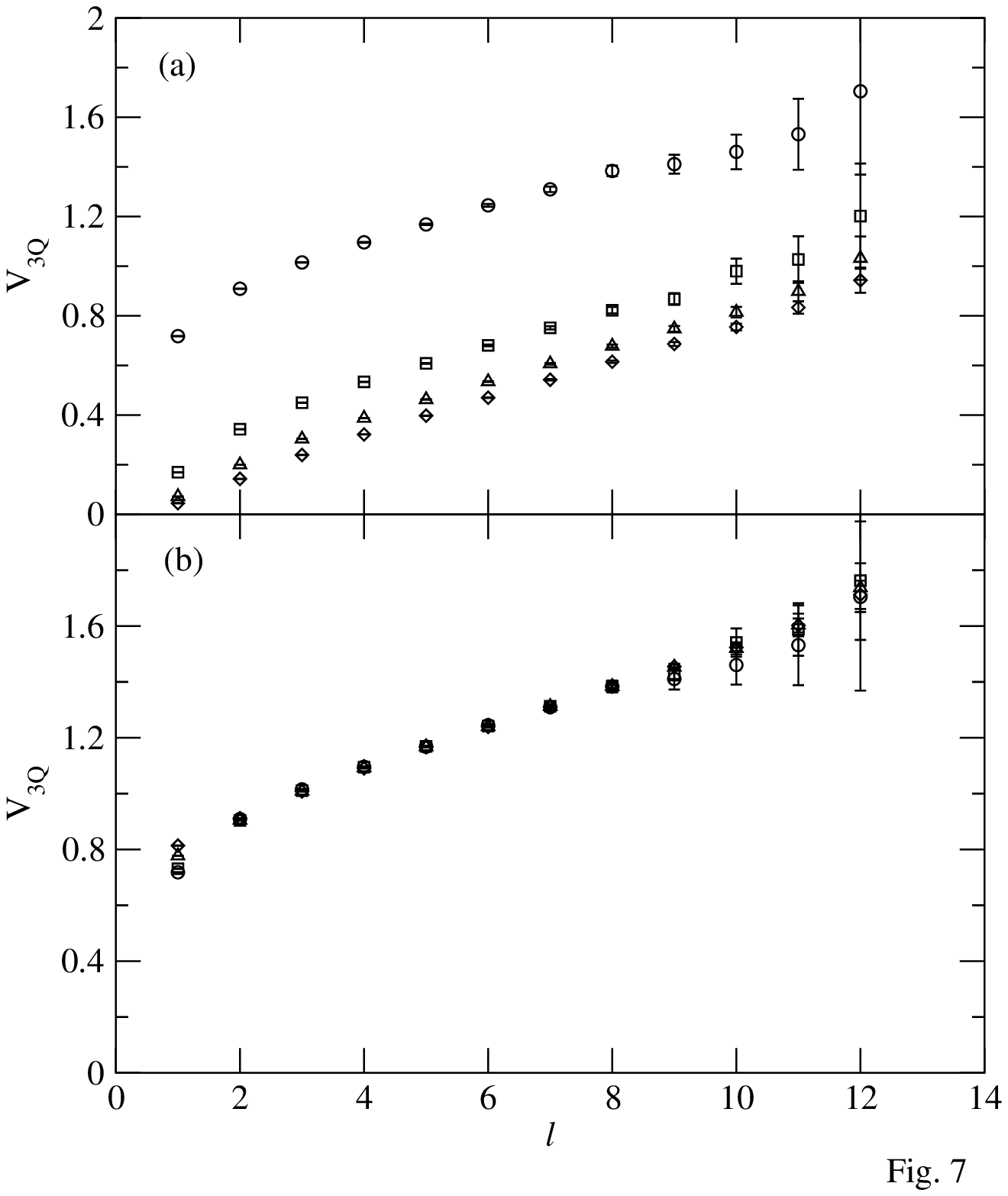}
\end{figure}
\end{center}
\vspace*{\fill}

\pagestyle{empty}
\begin{figure}
\includegraphics{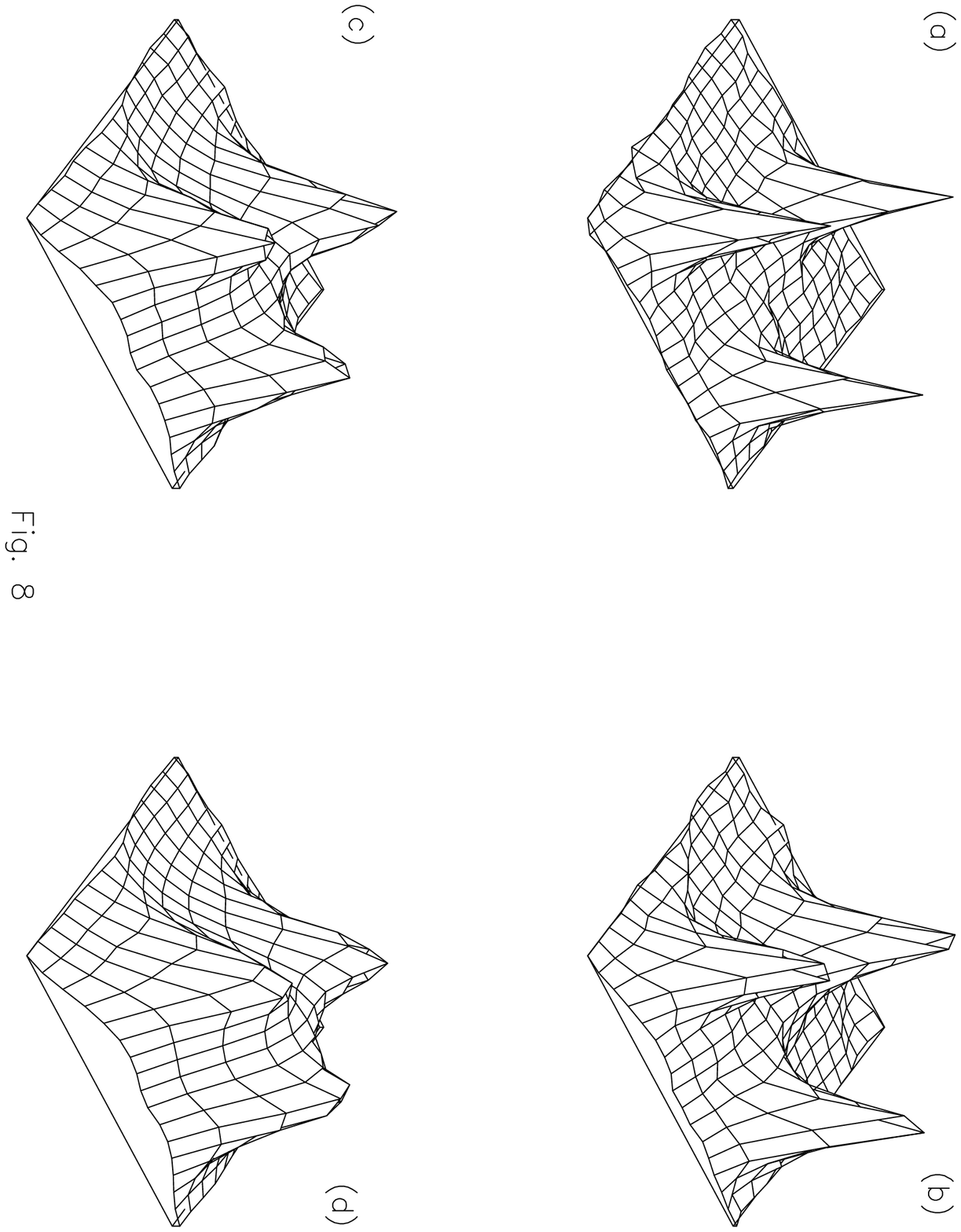}
\end{figure}

\pagestyle{empty}
\vspace*{\fill}
\begin{center}
\begin{figure}
\includegraphics{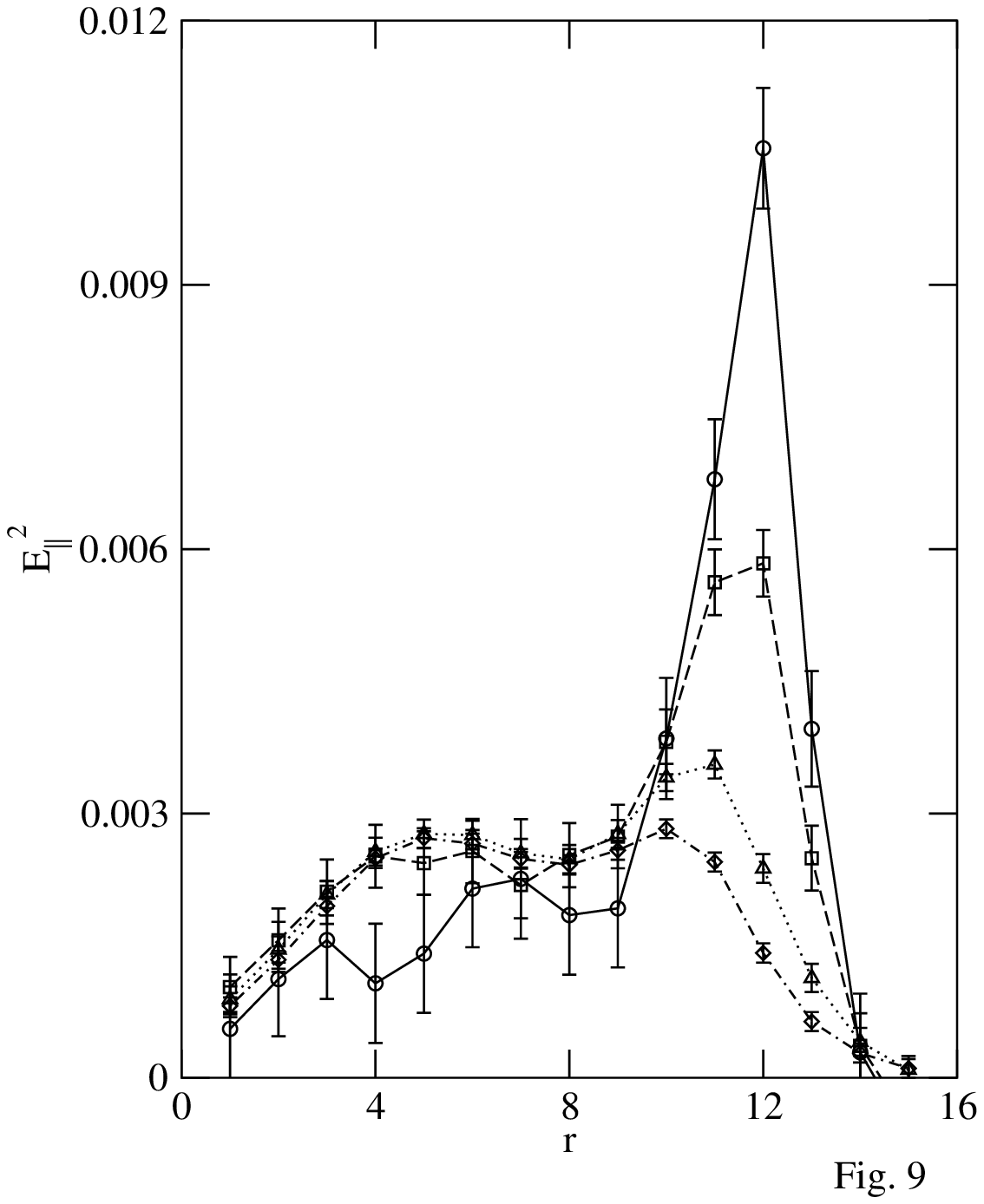}
\end{figure}
\end{center}
\vspace*{\fill}


\begin{thebibliography}{99}
\bibitem{parisi}G.~Parisi, R.~Petronzio, F.~Rapuano, 
Phys. Lett. B {\bf 128}, 418 (1983)

\bibitem{ForRoi}Ph.~de~Forcrand, C. Roiesnel, 
Phys. Lett. B {\bf 151}, 77 (1985)

\bibitem{ichie}H.~Ichie, V.~Bornyakov, T.~Streuer, G.~Schierholz,
Nucl. Phys. Proc. Suppl. {\bf 119}, 751 (2003)

\bibitem{lepage}G.P.~Lepage, 
Phys. Rev. D {\bf 59}, 074502 (1999)

\bibitem{orginos}K.~Orginos, D.~Toussaint, R.L.~Sugar, 
Phys. Rev. D {\bf 60}, 054503 (1999)

\bibitem{blum}T.~Blum, C.~DeTar, S.~Gottlieb, K.~Rummukainen, U.M.~Heller,
J.E.~Hetrick, D.~Toussaint, R.L.~Sugar, M.~Wingate,
Phys. Rev. D {\bf 55}, 1133 (1997)

\bibitem{hasen}S.~Hasenfratz, F.~Knechtli, 
Phys. Rev. D {\bf 64}, 034504 (2001)

\bibitem{folla}HPQCD Collaboration, E.~Follana {\it et al.}, 
hep-lat/0311004

\bibitem{wlee}W.~Lee, 
Phys. Rev. D {\bf 66}, 114504 (2002)

\bibitem{ape}APE Collaboration, M.~Albanese {\it et al.}, 
Phys. Lett. B {\bf 192}, 103 (1987)

\bibitem{bali}G.~Bali, K.~Schilling, Ch.~Schlichter, 
Phys. Rev. D {\bf 51}, 5165 (1995)

\bibitem{hoek}J.~Hoek, 
Nucl. Phys. B {\bf 329}, 240 (1990)

\bibitem{takahash}T.T.~Takahasi, H.~Matsufuru, Y.~Nemoto, H.~Suganuma, 
Phys. Rev. Lett. {\bf 65}, 18 (2001)

\bibitem{alex}C.~Alexandrou, Ph.~de~Forcrand, A.~Tsapalis, 
Phys. Rev. D {\bf 65}, 054503 (2002)

\bibitem{okiha}F.~Okiharu, R.M.~Woloshyn, 
hep-lat/0310007

\end{thebibliography}
\end{document}